\documentclass[aps,prb,twocolumn,showpacs,amsfonts,superscriptaddress,floatfix]{revtex4}

\usepackage{amsmath}
\usepackage{graphicx}
\usepackage{multimedia}
\usepackage{color}
\usepackage{ bbold }
\usepackage{bm}
\usepackage{float}

\newcommand{\beq}{\begin{equation}}
\newcommand{\eeq}{\end{equation}}
\newcommand{\beqarray}{\begin{eqnarray}}
\newcommand{\eeqarray}{\end{eqnarray}}

\newcommand{\Hc}{\ensuremath{\mbox{H.c.}}} 
\begin{document}

\title{Supersolid phases of light in extended Jaynes-Cummings-Hubbard systems} 
\author{B. Bujnowski} 
\affiliation{School of Physics, University of Melbourne, Victoria 3010, Australia}
\author{J.K. Corso} 
\affiliation{School of Physics, University of Melbourne, Victoria 3010, Australia}
\author{A.L.C. Hayward} 
\affiliation{School of Physics, University of Melbourne, Victoria 3010, Australia}
\author{J.H. Cole}
\affiliation{Chemical and Quantum Physics, School of Applied Sciences, RMIT University, Victoria 3001, Australia}
\author{A.M. Martin}
\affiliation{School of Physics, University of Melbourne, Victoria 3010, Australia}

\date{\today}

\begin{abstract}
Jaynes-Cummings-Hubbard lattices provide unique properties for the study of correlated phases as they exhibit convenient state
preparation and measurement, as well as {\it in situ} tuning of parameters. We show how to realize charge density and
supersolid phases in Jaynes-Cummings-Hubbard lattices in the presence of long-range interactions. The long-range interactions are realized by the
consideration of Rydberg states  in coupled atom-cavity systems and the introduction of additional capacitive couplings in
quantum-electrodynamics circuits. We demonstrate the emergence of supersolid and checkerboard solid phases, for calculations which take into account nearest neighbour couplings,  through a mean-field decoupling.
\end{abstract}


\maketitle
\section{Introduction}
Systems in a supersolid phase possess a spontaneously formed crystalline structure along with off diagonal long-range order which
characterizes superfluidity.
The investigation of supersolid phases in condensed matter systems has been a focus of research for more than half a century~\cite{Phys.Rev.106.161.1957,Sov.Phys.JETP.29.1107.1969,Ann.Phys.52.403.1969,Phys.Rev.Lett.25.1543.1970,PhysRevA2.256.1970}.
Until recently this effort has primarily focused on possible realization of a supersolid phase in $^4$He~\cite{J.Low.Temp.Phys.168.221.2012,J.Low.Temp.Phys.142.91.2006,J.Low.Temp.Phys.156.9.2009}, with the most credible claim for observation~\cite{Nature.427.225.2004}
now being withdrawn~\cite{Phys.Rev.Lett.109.155301.2012}.
The relatively recent realization of Bose-Einstein condensates, such as $^{52}$Cr~\cite{Phys.Rev.Lett.94.160401.2005,Phys.Rev.A.77.061601.2008},
$^{164}$Dy~\cite{Phys.Rev.Lett.107.190401.2011} and $^{168}$Er~\cite{Phys.Rev.Lett.108.210401.2012}, composed of atoms with large
dipole moments~\cite{Rep.Prog.Phys.72.126401.2009}, has provided an alternative avenue to investigate supersolid phases in extended
Bose-Hubbard lattice models~\cite{Phys.Rev.Lett.88.170406.2002,Europhys.Lett.72.162.2005,Phys.Rev.Lett.94.207202.2005,Phys.Rev.Lett.95.033003.2005,Phys.Rev.A.73.051601.2006,Phys.Rev.Lett.95.260401.2006}.

In this work we investigate the emergence of charge density wave and supersolid phases in Jaynes-Cummings-Hubbard (JCH) lattices.
Conventionally JCH lattices consist of an array of coupled cavities, with each cavity mode coupled to a
two-level system. A JCH system could be realised in, for example, photonic band-gap structures~\cite{Nat.Phys.2.849.2006,Nat.Phys.2.856.2006} and
coupled-cavity waveguides~\cite{PhysRevA.76.031805.2007, PhysRevLett.99.186401.2007}, arrays of superconducting strip-line
cavities~\cite{NewJPhys.12.093031},
or micro-cavities with individual cold-atoms connected via optical fiber interconnects~\cite{ApplPhysLett.87.211106.2005,
NatPhys.1.23.2005, Science.316.1007.2007}.
To date JCH systems are predicted to exhibit a
number of solid state phenomena such as: superfluid and Mott insulator
phases~\cite{Nat.Phys.2.849.2006,Nat.Phys.2.856.2006,PhysRevA.76.031805.2007,PhysRevLett.103.086403.2009,PhysRevA.80.023811.2009,PhysRevLett.104.216402.2010,PhysRevA.81.061801.2010}, the Josephson
effect~\cite{OptExpress.18.14586.2010}, meta-material properties~\cite{OptExpress.19.11018.2011}, Bose-glass
phases~\cite{PhysRevLett.99.186401.2007} and
fractional quantum Hall physics~\cite{PhysRevLett.108.223602.2012}. The JCH model has recently been experimentally realised for two sites using the internal and radial phonon states of two trapped ions \cite{PhysRevLett.111.160501.2013}.

Through the inclusion of a long-range interaction between the two-level systems we show that it is possible for supersolid phases to
emerge in the JCH-Hamiltonian.
To enable the long-range interaction we consider two cases: i) coupled microcavities with a single atom in each cavity and ii)
arrays of superconducting strip-line cavities. For microcavities the long-range interaction is achieved by accessing Rydberg states inducing a dipole
interaction between the atoms in each cavity. For arrays of superconducting strip-line cavities the long-range interaction is
mediated via capacitive couplings within the circuit.

\begin{figure}
\includegraphics[width=\columnwidth]{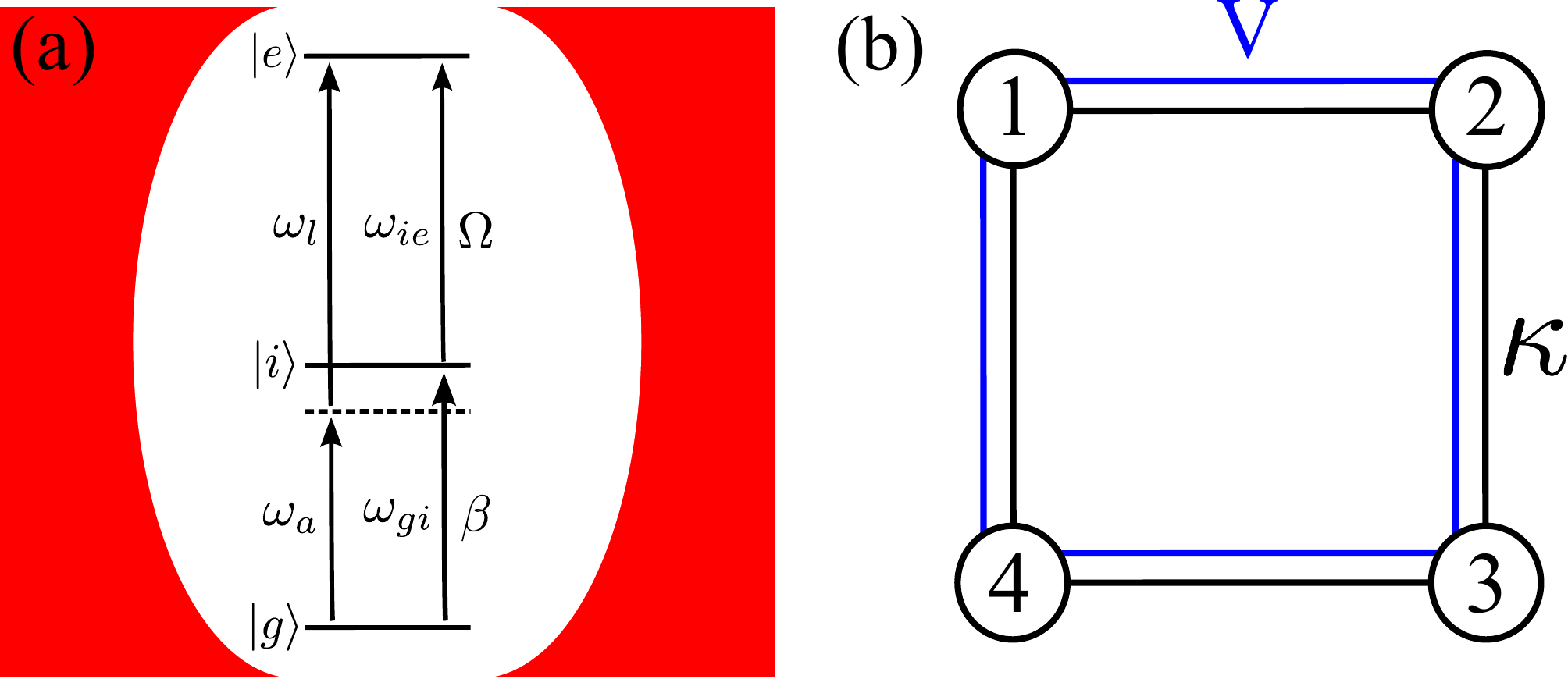}
   \vspace{-0.5cm}
   \caption{(Color online) (a) Scheme of a photon cavity containing a three level atom. The ground state $|g\rangle$ is resonantly coupled to the Rydberg state
   $|e\rangle$ via a two photon process. First, the atom is excited non resonantly by a photon of the cavity mode with frequency
   $\omega_a$. The transition to the Rydberg level happens by absorbing a photon from a driving laser field with frequency, $\omega_l$. b) Block of four cavities and with nearest-neighbour photon hopping, $\kappa$, (black lines) and nearest-neighbour interactions, $V$,  (blue lines).
   }\label{rydberg}
   \vspace{-0.5cm}
\end{figure}

Previous work has shown that, for a lattice of Rydberg atoms within a single cavity a supersolid phase can
emerge, where both superradiance and crystaline orders coexist~\cite{PhysRevLett.110.090402.2013}. In addition, driven coupled cavity systems
are also predicted to result in a supersolid phase~\cite{arxiv.1302.2242v1.2013}.

In this paper we initially (Section II) focus on coupling to Rydberg states in a single cavity containing a single atom, via a two photon process. In Section III we introduce coupling between the atom cavities, mediated via both photon tunnelling, between cavities, and dipole-dipole interactions, between atoms. In Section IV we then consider exact solutions for a system of four coupled atom cavities, specifically focusing on the role of nearest neighbour interactions. We then consider, in Section V, mean-field solutions, demonstrating the emergence of checkerboard solid and supersolid phases in the presence of nearest neighbour interactions.   
 
 \section{Two Photon Coupling to Rydberg States in a single atom cavity}
 
Before considering a lattice we focus on the properties of a single site. To achieve long-range interactions in the coupled
atom-cavity system we require the excited state of the atom to have a significant dipole moment. A possible realization of an atomic
cavity that exhibits a dipole moment when excited, utilizes the Rydberg state of $^{87}$Rb atoms.
The $5S_{1/2}$ ground state $|g\rangle$ of the $^{87}$Rb atom, that has been placed inside the cavity,
is resonantly coupled to the Rydberg state $|e\rangle$ via a two photon process, by using the $5P_{3/2}$ state $|i\rangle$ as an
intermediate step [Fig.~1(a))]. By choosing appropriate detunings for the driving fields the intermediate state can be eliminated adiabatically
as there are only small changes in its population over time. 
As schematically shown in Fig.~1(a) the transition from the ground state $|g\rangle$ to intermediate state $|i\rangle$ of frequency $\omega_{gi}$ is driven by the non resonant coupling of strength $\beta$ to a single quantized cavity mode of frequency $\omega_a$ and photonic annihilation operator
$\hat{a}^\dagger$, where the cavity mode is detuned by $\Delta=\omega_a-\omega_{gi}$. The remaining transition from $|i\rangle$ to 
$|e\rangle$ of frequency $\omega_{ie}$ is non resonantly driven by a classical laser field with frequency $\omega_l$ and Rabi frequency
$\Omega$. Transferring the interacting part of the Hamiltonian into the interaction picture and applying the rotating wave approximation, the
Hamiltonian can be written in the following form ($\hbar=1$)
\begin{align}\label{Ryd2}
      \hat{\mathcal{H}}^{\mathrm{int}}_1&=-\Delta|i\rangle\langle i|+\left(\beta\hat{a}^\dagger|g\rangle\langle
      i|+\Omega|e\rangle\langle i|+\Hc\right).
\end{align}
For large detuning $\Delta$ compared to the lifetime of $|i\rangle$ the intermediate state is weakly populated and can be eliminated adiabatically.
This leads to the effective Hamiltonian 
\begin{align}\label{Ryd3}
      \hat{\mathcal{H}}^{\mathrm{eff}}&=\tilde{\beta}\left(\hat{a}^\dagger|g\rangle\langle e|+\hat{a}|e\rangle\langle g|\right),
\end{align}
where $\tilde{\beta}$ denotes the rescaled coupling strength between the cavity mode and the atom. 
Thus the three level atomic system is approximated by a two-level system, with the excited state of the atom exhibiting a
significant dipole moment. For a detailed description of the two cavity system, including experimentally accessible parameter regimes see the work of Wu {\it et al.} \cite{PhysRevA.88.043816.2013}.

An alternative system in which to realise the Jaynes-Cummings Hubbard model is circuit quantum-electrodynamics (cQED)~\cite{Nature.431.162.2004,Nature.445.515.2007,PhysRevA.69.062320,Nat.Comm.4.1420,arXiv:1312.2963,AnnalendePhysik.525.395}. In a typical QED circuit, the `atomic' degree of
freedom is realised via a Josephson junction circuit providing a non-linear set of states, the lowest two of which form
an effective two level system. The photonic degree of freedom is formed from the quantized modes of a superconducting
strip-line resonator.  The resulting system Hamiltonian takes on an equivalent form to that of Eq.~\eqref{Ryd3}.  This system
provides several advantages, including strong atom-photon coupling and ease of integration as the form of the effective
Hamiltonian can be tailored at the circuit design stage. 

 \section{The extended Jaynes-Cummings-Hubbard Model}
 \begin{figure}
 \includegraphics[width=1.1\columnwidth]{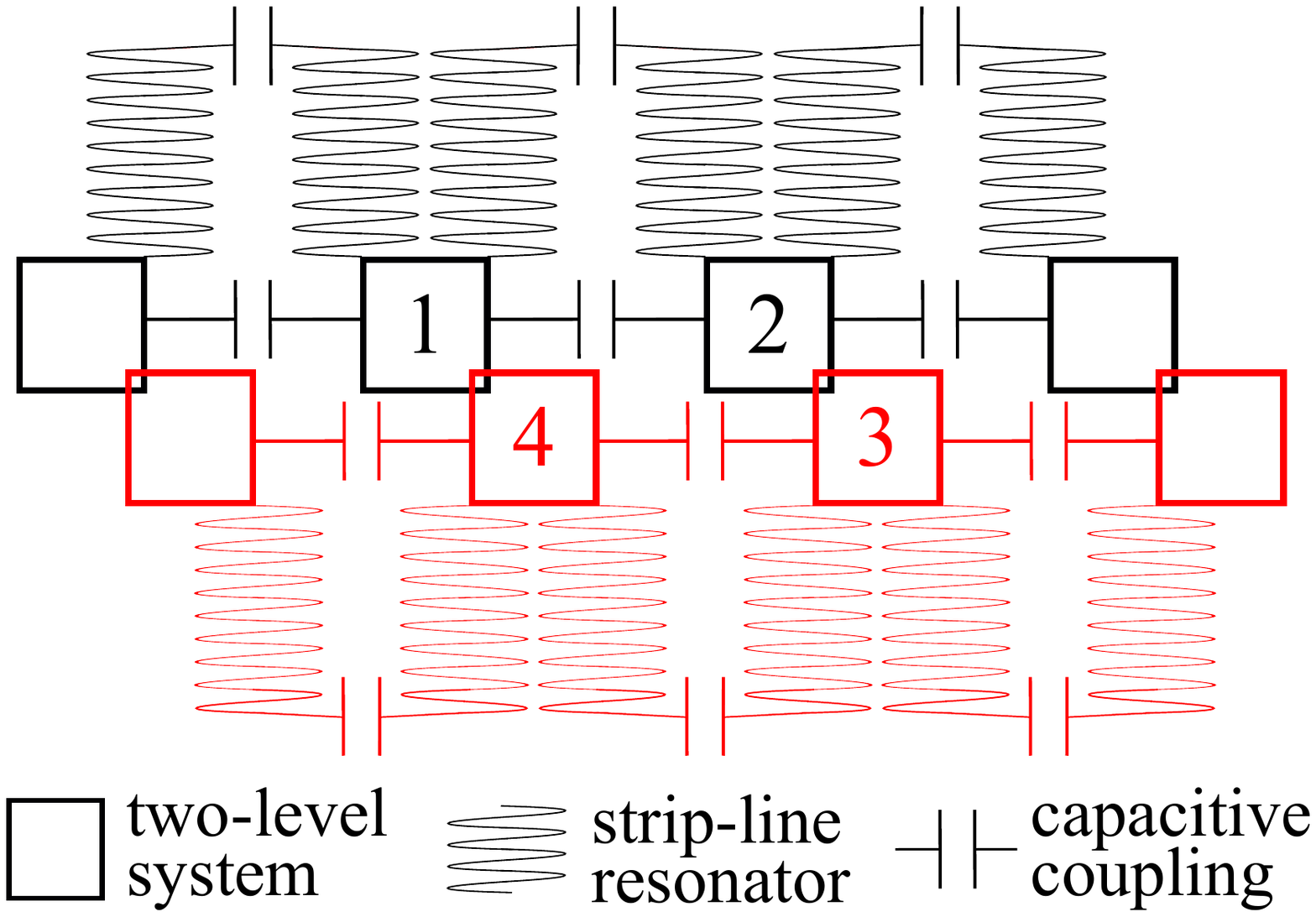}
   \vspace{-8.4cm}
   \caption{(Color online) Schematic of a possible cQED realization of the long-range JCH model. Here we show two layers (black and red) of a multi-layered circuit. In each layer Josephson junction based two-level systems are coupled via strip-line resonators and capacitors, as denoted in the figure. Multiple layers of these one dimensional arrays are placed on top of each other to realize an effective 2-D lattice.
   }\label{Cqed}
 \vspace{-0.5cm}
\end{figure}
\begin{figure*}
\includegraphics[width=2.1\columnwidth]{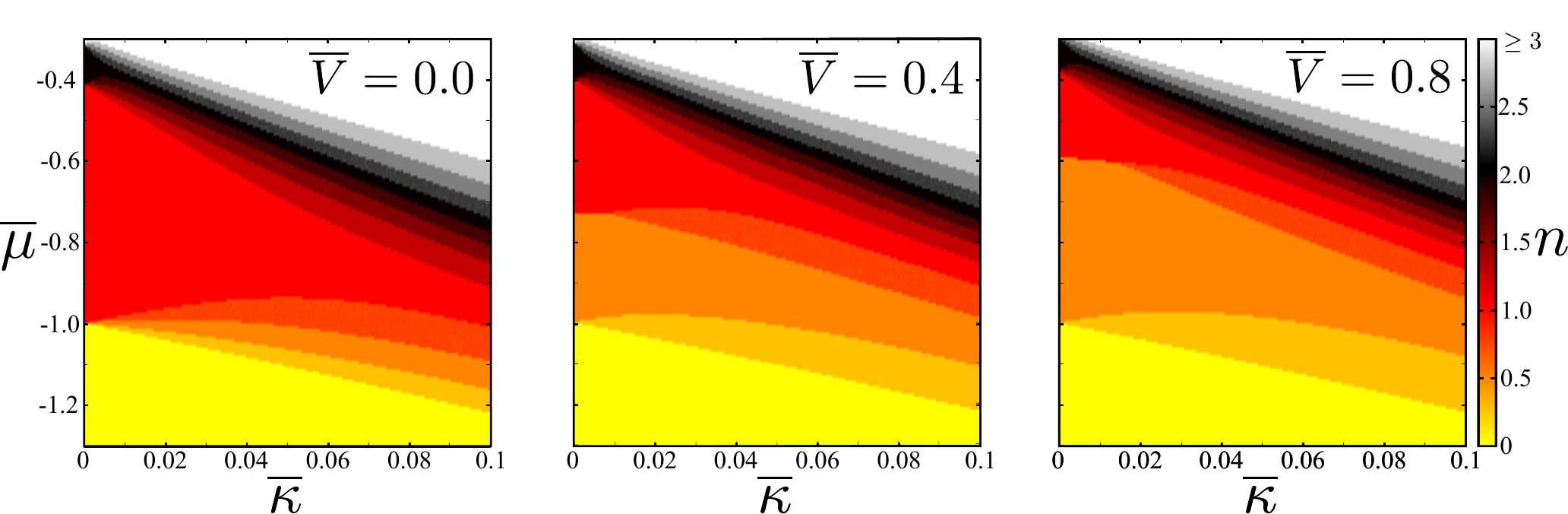}
   \vspace{-0.5cm}
\caption{(Color online) Number of excitations per site, $n=\frac{1}{4}\sum_{i=1}^4\langle \hat{l}_i\rangle$, in a system of four coupled atom cavities, with periodic boundary conditions, as a function of the chemical potential $\bar{\mu}$ and the inter-cavity hopping $\bar{\kappa}$, for different values of the nearest neighbour interaction
$\bar{V}$, with $\bar{\Delta}=0$.}
       \label{exact}
        \vspace{-0.5cm}
\end{figure*}

An array of coupled photon cavities that contain two-level atoms can be described by the
JCH model~\cite{Nat.Phys.2.849.2006,Nat.Phys.2.856.2006,PhysRevA.76.031805.2007,PhysRevB.81.104303,PhysRevA.83.023821,Comp.Phys.Comm.182.2036}. In this paper we extend the JCH to the
case of Rydberg atoms by including a dipole interaction term. In the grand canonical ensemble the Hamiltonian
for this system is defined as
\begin{align}\label{eJCH}
      \hat{\mathcal{H}}&=-\kappa\sum_{\langle  i, j \rangle}\hat{a}_i^\dagger \hat{a}_j
      +\tilde{\beta}\sum_i\left( \hat{a}^\dagger_i \hat{\sigma}_i+\hat{a}_i \hat{\sigma}^\dagger_i \right)
      +\sum_i\left(\epsilon\hat{n}_i^\sigma+\omega \hat{n}_i^a\right)\notag\\
      &+\frac{V}{2}\sum_i\sum_{j \neq
      i}\frac{\hat{n}_i^\sigma\hat{n}_j^\sigma}{\left|\mathbf{r}_i-\mathbf{r}_j\right|^3}-\mu\sum_i\hat{l}_i.
\end{align}
The above Hamiltonian does not include dissipation, driving and other non-equilibrium effects, which will be present in experiment, however it does provide a simple, equilibrium, model which can be used to investigate the possibility of new phases in such systems. The first term describes the hopping of photons between neighbouring lattice sites $i$ and $j$ with hopping amplitude $\kappa$,
where $\hat{a}_i^\dagger$ and $\hat{a}_i$ are photon creation and annihilation operators at lattice site $i$.
The second term is the on site coupling between the photons and atoms for each site $i$ as in
Eq.(\ref{Ryd3}), where $\hat{\sigma}^\dagger_i=|e_i\rangle\langle g_i|$ and $\hat{\sigma}_i=|g_i\rangle\langle e_i|$ are atomic
raising and lowering operators on site $i$ respectively. The next term defines the energy associated with the atomic and photonic
degrees of freedom on each site, where $\epsilon$ is the energy of the Rydberg level and $\omega$ is the frequency of the photon
mode of the cavity.
Here we define the atom number operator $\hat{n}_i^\sigma=\hat{\sigma}^\dagger_i\hat{\sigma}_i$ that counts the atoms in the
excited state and the photon number operator $\hat{n}_i^a=\hat{a}_i^\dagger\hat{a}_i$. The fourth term is the dipole-dipole interaction between the atoms
with interaction strength $V$. It raises the total energy of the system if atoms in two or more different cavities occupy excited states.
The JCH model does not conserve the number of photonic or atomic excitations. However the total number of excitations $\sum_i
\hat{l}_i=\sum_i \left(\hat{n}_i^\sigma+\hat{n}_i^a\right)$ is conserved. The last term in Eq.~\eqref{eJCH} specifies the total number
of excitations in the system via the chemical potential $\mu$.

For the cQED case, arrays of coupled cavities can be fabricated with either capacitive or inductive coupling linking the resonators.
This architecture provides an entirely equivalent realization of the JCH model~\cite{NewJPhys.12.093031,NewJPhys.14.075024}.
In principle, coupling the `atoms' in a cQED system can be achieved via direct qubit-qubit
coupling~\cite{RMP} or via the inclusion of  additional components to provide an effective coupling
term~\cite{PhysRevB.74.140504,PhysRevB.76.174507,Phys.Rev.B.78.064503,PhysRevB.79.020507,PhysRevB.77.014510}. Although the
exact functional form of the coupling will depend strongly on the particular circuit realisation, Fig.~2 provides a schematic
for two layers of a possible multiple layer circuit. Each layer in the circuit consists of a one dimensional array of Josephson
junction based two-level systems coupled via strip-line resonators and capacitors. In such a circuit the photonic components of the
JCH model are now microwave excitations in the strip-line resonators and the long-range interactions (in this case
nearest-neighbour) arise from capacitive coupling between adjacent Josephson junction two-level systems. For a multi-layered system
capacitive coupling between strip-line resonators in adjacent layers enables microwave excitations to couple between layers.
Additionally, capacitive coupling between Josephson junctions in adjacent layers mediates a long-range interaction between two-level systems. Multiple
layers where coupling between strip-lines and Josephson junctions is only between adjacent layers and nearest neighbours form an
effective two-dimensional lattice. A useful variant is to use `flux-qubits' and LC-resonators for
the atomic and photonic components respectively, as this results in a smaller circuit footprint~\cite{Nature02831,PhysRevB.78.180502}.

For the coupled atom cavity system the dipole interactions decay as $\left| \mathbf{r}_i -\mathbf{r}_j\right|^{-3}$, see second to last term in Eq. (\ref{eJCH}). For the cQED case the exact parameterisation of the of the long-range interactions depends on the capacitive couplings between qubits, with the exact coupling defined by the circuit geometry. The aim of this work is to show that for either a coupled atom cavity system or the cQED case extended interactions will lead to new charge density wave and supersolid phases. As such we consider the simplest form of extended interactions (nearest neighbour), with the resulting Hamiltonian:
\begin{align}\label{eJCH_nn}
      \hat{\mathcal{H}}&=-\kappa\sum_{\langle  i, j \rangle}\hat{a}_i^\dagger \hat{a}_j
      +\tilde{\beta}\sum_i\left( \hat{a}^\dagger_i \hat{\sigma}_i+\hat{a}_i \hat{\sigma}^\dagger_i \right)
      +\sum_i\left(\epsilon\hat{n}_i^\sigma+\omega \hat{n}_i^a\right)\notag\\
      &+\frac{V}{2}\sum_{\langle  i, j \rangle}\hat{n}_i^\sigma\hat{n}_j^\sigma-\mu\sum_i\hat{l}_i.
\end{align}

\section{Four site solutions of the extended Jaynes-Cummings-Hubbard Model}
In Section V we will consider mean-field solutions for an infinite array of coupled atom cavities. The mean-field approximation will be based on the coupling of a four site system to an infinite lattice. As such, before proceeding with the mean-field coupling  we first consider the exact solutions of a four site system. The  system under consideration is  schematically shown in Fig.~1(b), i.e. four atom cavities in a square arrangement, with nearest-neighbour hopping and nearest-neighbour interactions. For such a system the Hamiltonian is 
\begin{eqnarray}
 \hat{\mathcal{H}}_{4}&=&-z\bar{\kappa}\left(\hat{a}_1^\dagger \hat{a}_2+\hat{a}_2^\dagger \hat{a}_1 +\hat{a}_2^\dagger \hat{a}_3+\hat{a}_3^\dagger \hat{a}_2 \right. \nonumber \\ &+& 
 \left.\hat{a}_3^\dagger \hat{a}_4+\hat{a}_4^\dagger \hat{a}_3+\hat{a}_1^\dagger \hat{a}_4+\hat{a}_4^\dagger \hat{a}_1\right) \nonumber \\
 &+& \sum_{i=1}^4\left(-\bar{\mu} \hat{n}_i^a -\left(\bar{\Delta}+\bar{\mu}
      \right)\hat{n}_i^\sigma+\left(\hat{\sigma}_i^\dagger\hat{a}_i+\hat{a}^\dagger\hat{\sigma}_i\right)\right)
 \nonumber \\
 &+&z\bar{V}\left(\hat{n}_1^{\sigma}\hat{n}_2^{\sigma}+\hat{n}_2^{\sigma}\hat{n}_3^{\sigma}+\hat{n}_3^{\sigma}\hat{n}_4^{\sigma}+\hat{n}_4^{\sigma}\hat{n}_1^{\sigma}\right),
 \label{4site}
\end{eqnarray}
where $z=1$ ($z=2$) for fixed (periodic) boundary conditions. In the above we have introduced the following dimensionless couplings: $\bar{\kappa}=\kappa/\tilde{\beta}$, $\bar{\mu}=(\mu-\omega)/\tilde{\beta}$, $\bar{\Delta}=(\omega-\epsilon)/\tilde{\beta}$ and $\bar{V}=V/(\tilde{\beta}l_0)$, where $l_0$
is the lattice unit length. Note that switching between periodic and fixed boundary conditions can be included by a trivial rescaling of ${\overline \kappa}$ and ${\overline V}$.

To find the ground state of the four site system we evaluate Eq. (\ref{4site}), in the following basis: $ \left\{\prod_{i=1}^4 |n_i^a,n_i^\sigma\rangle; n_i^a\in\mathbb{N}_0,n_i^\sigma\in\left\{0,1\right\} \right\}$. Due to computational limitations it is necessary to restrict the possible basis states. In the following a cut-off defined by  $n_i^a+n_i^\sigma\leq5$ was used on every site, which is more than sufficient to demonstrate the fundamental ground state properties of the four site system. 

Fig. 3 plots the number of excitations per site ($n=\frac{1}{4}\sum_{i=1}^4\langle \hat{l}_i\rangle$), in the ground state, for the four site system, with periodic boundary conditions ($z=2$) as a function of the chemical potential and photon hopping strength, for various strengths of nearest neighbour interactions, with $\bar{\Delta}=0$. In the absence of dipolar interactions ($\bar{V}=0$) we observe a {\it pinch} effect as ${\overline \kappa} \rightarrow 0$ between $n = k$ and $n = k+1$, where $k=1,2,3,..$, i.e. all fractional occupations disappear as  ${\overline \kappa} \rightarrow 0$, consistent with previous results. This implies that in the limit of no photon hopping between the cavities the ground state is defined by the number of excitations being an integer multiple of the number of sites. As photon hopping is introduced states emerge with the number of excitations not being  an integer multiple of the number of sites. Introducing nearest neighbour interactions significantly changes the ground state properties of the system. Specifically, we see that as ${\overline \kappa} \rightarrow 0$ fractional occupations do not disappear, i.e. for ${\overline \kappa} \rightarrow 0$ a new state emerges, with $n=0.5$. This corresponds to a checkerboard phase with atomic excitations arranged on the diagonal to minimise the nearest neighbour interaction. As the strength of the nearest neighbour interaction increases the chemical potential range over which this fractional state exists grows. For the fractional $n=0.5$ state, in the limit ${\overline \kappa} \rightarrow 0$, we find the the atomic and photonic contributions to this state are equal, i.e. $\sum_{i=1}^4\langle {\hat n}_i^{\sigma}\rangle= \sum_{i=1}^4\langle {\hat n}_i^{a}\rangle =1$. Additionally,  for the $n=1$ and $n=2$ states  the introduction of nearest neighbour interactions increases (decreases) the photonic (atomic) contributions to the number of excitations per site. The emergence of the $n=0.5$ checkerboard state corresponds to a checkerboard solid. The increase (decrease) in the photonic (atomic) contributions to the number of excitations per site (for $n=1$ and $n=2$)  comes from the energy cost of having atomic excitations in the system, due to the nearest neighbour interactions.

 \section{Meanfield solutions of the extended Jaynes-Cummings-Hubbard Model}

\begin{figure}
 \includegraphics[width=1.1\columnwidth]{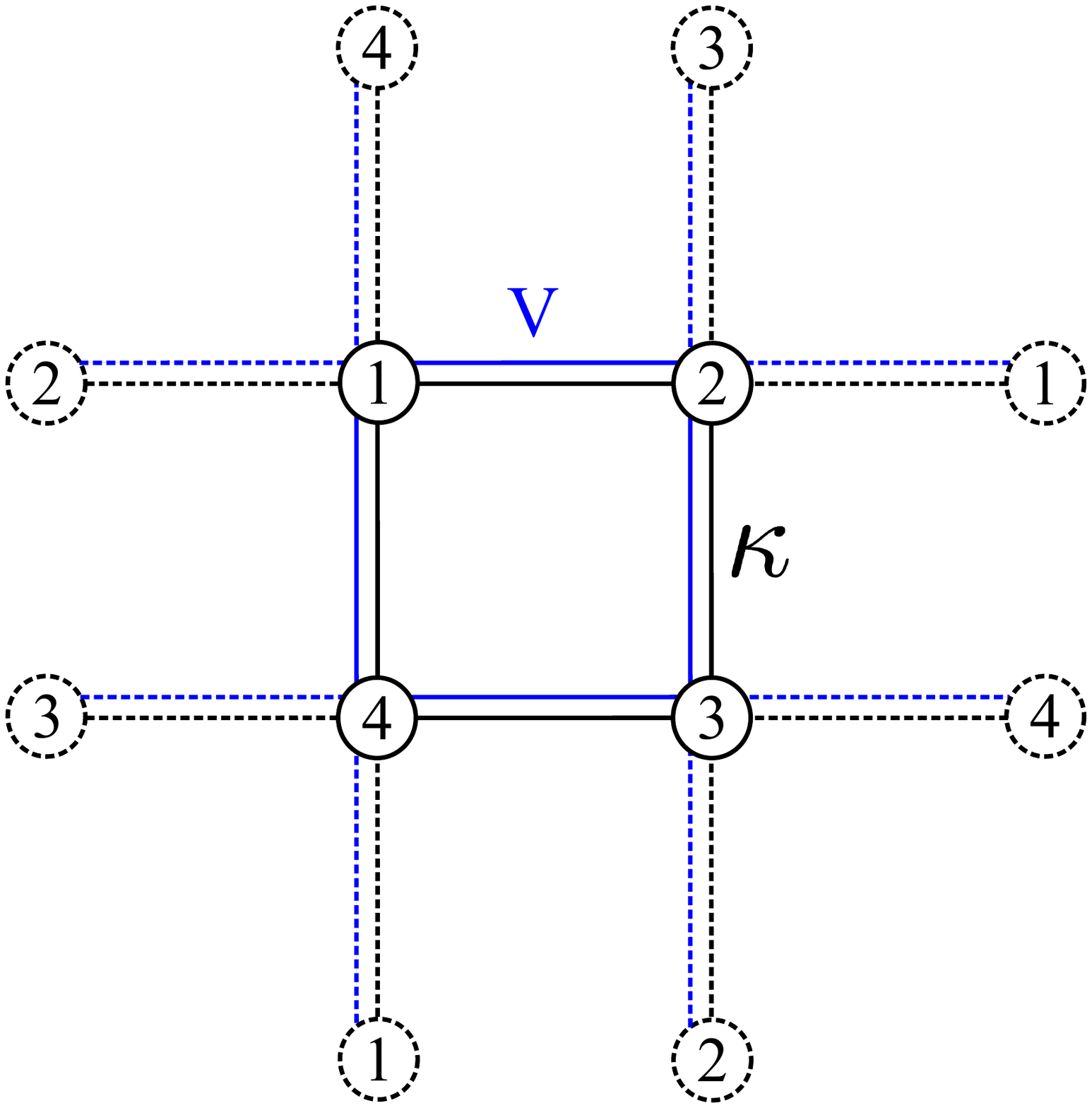}
   \vspace{-6.0cm}
   \caption{(Color online) Scheme of the mean-field approximation for a block of four cavities and nearest neighbour interactions. The intra block hopping, $\kappa$, (black solid line) and nearest-neighbour dipole interactions, $V$, (blue solid line) between neighbouring cavities (black circles) are treated exactly. Interactions outside of the block are decoupled by a mean-field approximation (black and blue dashed lines).
   }\label{Mean_field}
   \vspace{-0.5cm}
\end{figure}

\begin{figure*}
\includegraphics[width=2.1\columnwidth]{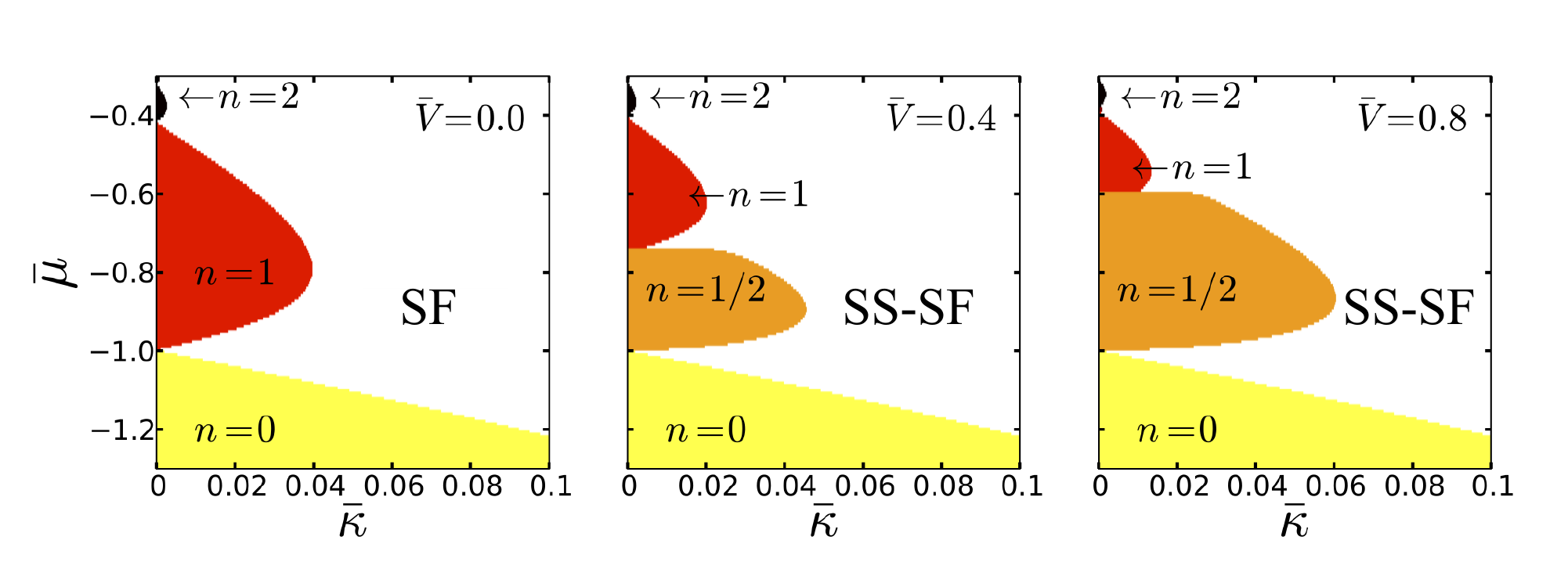}
   \vspace{-0.5cm}
\caption{(Color online) Mean number of excitations per block of four cavities $n=\frac{1}{4}\sum_{i=1}^4\langle \hat{l}_i\rangle$ within the Mott insulator and checkerboard phases, as a function of the chemical potential $\bar{\mu}$ and the inter-cavity hopping $\bar{\kappa}$, for different values of the nearest neighbour interaction
$\bar{V}$, with $\bar{\Delta}=0$. In the white SS-SF region we find non vanishing superfluid order parameters, i.e. $\sum_{i=1}^4 \psi_i>0$.}
       \label{phase1}
        \vspace{-0.5cm}
\end{figure*}

\begin{figure}
\includegraphics[width=\columnwidth]{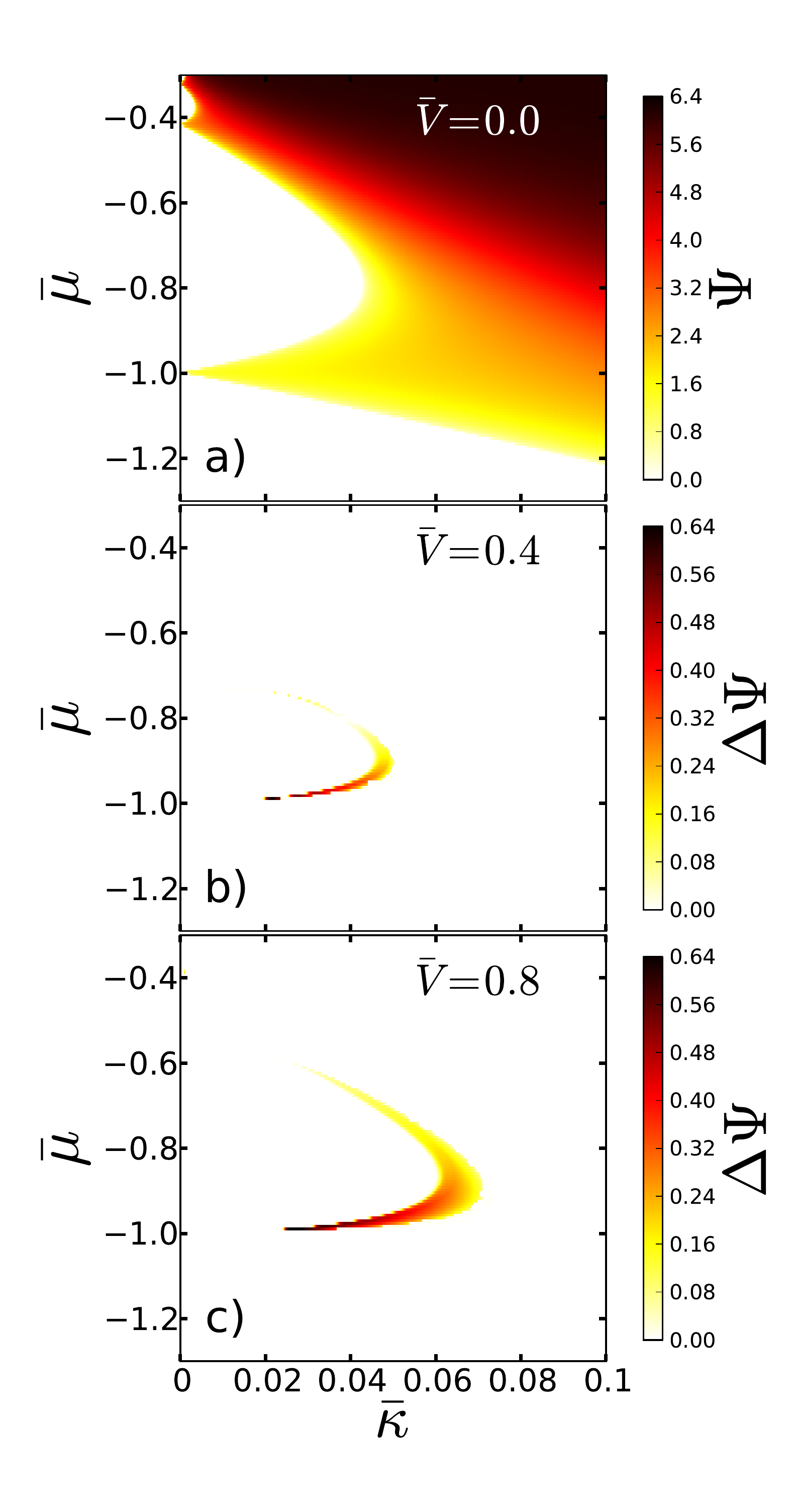}
   \vspace{-1.2cm}
\caption{(Color online) a) The sum of the superfluid order parameters on each block $\Psi=\sum_{i=1}^4\psi_i$ as a function of the chemical
potential $\bar{\mu}$ and the inter-cavity hopping $\bar{\kappa}$ for $\bar{V}=0.0$. b) and c) $\Delta\Psi=|\psi_1-\psi_2|/|\psi_1+\psi_2|$ as a function 
of the chemical potential $\bar{\mu}$ and the inter-cavity hopping $\bar{\kappa}$ for $\bar{V}=0.4$ and $0.8$. This normalized difference between the
order parameters indicates the presence of a checkerboard super solid (SS) phase.}\label{phase2}
 \vspace{-0.6cm}
\end{figure}

For the case of vanishing long-range interaction ($\bar{V}=0$) phase transitions between homogeneous superfluid and homogeneous Mott-insulator
phases~\cite{Nat.Phys.2.849.2006,Nat.Phys.2.856.2006,PhysRevA.76.031805.2007,PhysRevLett.103.086403.2009,PhysRevA.80.023811.2009} have been predicted.
To determine these phases one approach is to introduce a mean-field decoupling in the hopping term,
between the lattice sites. For a single site decoupling this restricts the excitations to be homogeneously distributed.

Including the nearest neighbour interactions ($\bar{V}\neq0$) we expect a non-homogeneous distribution of excitations due to the energy cost of having more
than one atom in the excited state close to each other. To observe this in the mean-field model, we allow the order parameters to vary across the system.
For interactions between nearest neighbours only, variations of the order parameter appear
with a maximum period of two lattice unit lengths. Thus it is sufficient to decouple the infinite square lattice of cavities into
periodical square blocks of four, as depicted in Fig.~\ref{Mean_field}. Tunneling (dashed lines) and long-range interactions between
blocks (doted lines) are treated by a mean-field approximation, i.e. operator products that connect two blocks are approximated
by $\hat{A}\hat{B}\approx\hat{A}\langle{\hat{B}}\rangle+\langle{\hat{A}}\rangle\hat{B}-\langle\hat{A}\rangle\langle\hat{B}\rangle$.
Intra-block tunnelling and long-range interactions (solid lines) are kept to capture the significant effects of correlations within
the block. The approximated Hamiltonian can be written as a sum of terms within the block and mean field terms connecting the block
to the surrounding lattice sites,
$\hat{\mathcal{H}}\approx\hat{\mathcal{H}}'=\hat{\mathcal{H}}_{4}+\hat{\mathcal{H}}_{MF}$
where $\hat{\mathcal{H}}_{4}$ is given by Eq. (\ref{4site}), with $z=1$, and  $\hat{\mathcal{H}}_{\mathrm{MF}}=\hat{\mathcal{H}}^{12}_{\mathrm{MF}}
      +\hat{\mathcal{H}}^{23}_{\mathrm{MF}}+\hat{\mathcal{H}}^{34}_{\mathrm{MF}}+\hat{\mathcal{H}}^{43}_{\mathrm{MF}}$ with

\begin{align}
\hat{\mathcal{H}}_{\mathrm{MF}}^{ij}&
      =-\bar{\kappa}\left \{
            \left(\hat{a}_i^\dagger+\hat{a}_i\right)\psi_j+\left(\hat{a}_j^\dagger+\hat{a}_j\right)\psi_i-2\psi_i\psi_j\right
      \}\notag\\ 
&+\bar{V} \left \{\hat{n}_i\xi_j+\hat{n}_j\xi_i-\xi_i\xi_j\right\}.
\end{align} 
Here we introduced the mean-field parameter of the atomic excitation on each site $\xi_i=\langle\hat{n}_i^\sigma\rangle$ and the
real superfluid order parameter $\psi_i=\langle\hat{a}_i\rangle$.

To find the ground state of $\hat{\mathcal{H}}'$, we calculate the order parameters self-consistently,
minimizing the total energy of the system simultaneously. To avoid convergence problems, due to
degenerate ground states in a non-homogeneous phase, the symmetry of the block is broken by introducing a random energy shift of
$\delta_i$ for each cavity. This energy shift~\cite{gugu}, in dimensionless units $\bar{\delta}_i=\delta_i/\beta$, is of order
$10^{-6}$.
As in Section IV the basis is restricted to  $n_i^a+n_i^\sigma\leq5$ on each site. 

The phase diagram for our system can be deduced from Figs.~\ref{phase1} and~\ref{phase2}.
For $\bar{V}=0$ we find the well known result that the parameter space is separated into two distinct phases. For low hopping strength $\bar{\kappa}$
we find lobes of vanishing superfluid order parameter, i.e. Mott insulating phases as shown in Fig.~\ref{phase2}a).
Each lobe corresponds to a state with an integer number of strongly localized excitations per site. This is shown in
Fig.~\ref{phase1} where we plot the mean number of excitations per site in the Mott insulator phase. For low chemical potential there are no excitations in the system. Raising the chemical potential the block is successively filled with one, two and more excitations per site. 
At sufficiently large $\bar{\kappa}$ the system undergoes a phase transition into a phase of finite superfluid order parameter
[Fig.~\ref{phase2}a)]. The excitations are homogeneously distributed and delocalized, i.e. the system is in a superfluid state.

Consistent with the finite size results (Fig.~\ref{exact}) Fig.~\ref{phase1} shows that as $\bar{V}$ is increased from zero a new phase emerges, with $n=1/2$ and zero supersolid/superfluid
component, corresponding to two excitations per block. This corresponds to a checkerboard solid phase with excitations
arranged on the diagonal to minimise the nearest neighbour interaction. As the strength of the nearest neighbour interactions are increased
the extent of this checkerboard solid phase increases. Checkerboard solid phases also appear at higher filling
($n=\frac{3}{2}$, $n=\frac{5}{2}$). However, for the values of $\bar{V}$ considered the extent of phases is very small.

Associated with the emergence of the checkerboard solid phase is supersolid behaviour. Supersolid regimes can be
characterized by identifying changes in the superfluid order parameters within the block. In the absence of nearest neighbour interactions, $\psi_1=\psi_2=\psi_3=\psi_4$. However, for $\bar{V}>0$ we find new regimes where $\psi_1=\psi_3$ and
$\psi_2=\psi_4$ but $\psi_1\neq\psi_2$ corresponding to a supersolid phase. Figures \ref{phase2}b) and c) identify the supersolid phase by plotting $\Delta\Psi=|\psi_1-\psi_2|/|\psi_1+\psi_2|$, in the region where  $\sum_{i=1}^4\psi_i>0$. As can be seen from Fig.~\ref{phase1} this
supersolid phase is present at the interface of the checkerboard solid phase and as $\bar{\kappa}$ is increased it diminishes, until $\psi_1=\psi_2$ characterizing a superfluid state. As for the checkerboard solid phase, as $\bar{V}$ is increased the extent of the supersolid phase increases.

Intra-block correlations play a crucial role in determining the phase diagram in the extended JCH model. Introducing mean-field decouplings
within the block for the hopping and nearest neighbour interactions [first and last terms in
Eq.(5)] changes the phase diagram for $\bar{V}\neq 0$. Decoupling results in a decrease of the extent of the checkerboard solid and supersolid phases. These shifts grow in significance when $\bar{V}\rightarrow 1$. This is in contrast to the case $\bar{V}=0$ where
the phase diagram for a single site decoupling~\cite{Nat.Phys.2.849.2006,Nat.Phys.2.856.2006,PhysRevA.76.031805.2007,PhysRevA.80.023811.2009}
and the block wise decoupling do not differ significantly. Additionally, for supersolid phases in the extended Bose-Hubbard model it has been found that quantum fluctuations can play a significant role in determining stability \cite{Phys.Rev.Lett.94.207202.2005}. As such to test the robustness of these mean-field calculations quantum Monte Carlo methods should be employed.

\section{Conclusions}

The nature of the nearest neighbour interaction in the JCH model is qualitatively different to that found in other lattice systems
with long-range interactions, such as ultra-cold dipolar gases in optical lattices, where the extended Bose-Hubbard model is
appropriate. Specifically in the extended JCH model the interaction is mediated via a two-level system. Thus the interaction depends on the
simultaneous excitation of neighbouring atoms which favors anti-ferromagnetic correlations between the atomic states. Indeed, at
$\kappa=0$ the JCH system maps to a quantum Heisenberg model \cite{PhysRevA.76.031805.2007,Phys.Rev.Lett.99.160501.2007,Phys.Rev.A.78.062338.2008,Euro.Phys.Lett.84.20001.2008}, in contrast to the Bose Hubbard case, which lies in the classical
Ising universality class.

We have demonstrated that the inclusion of long-range interactions in the JCH model results in the emergence of i) non-integer Mott
insulator phases and ii) supersolid phases. In absence of long-range interactions the two-level systems mediate an interaction
between photons in the lattice. The predicted Mott-insulator superfluid transition is a direct consequence of this interaction. 
The addition of a direct coupling between the two-level systems introduces charge density and supersolid phases.
Such an extended interaction can be mediated in coupled atom-cavity systems through the inclusion of Rydberg states and in cQED systems via
capacitive couplings between qubits. JCH systems provide an alternative platform to investigate the emergence of supersolid phases and novel correlated
states of light. 

Experimental realisations of the JCH model are subject to losses mechanisms, which are not included in the above work. As such the analysis presented above is valid for regimes where the quality factor of the cavities is large and the hopping rates between the cavities dominates over absorption/loss of photons out of the system. To study the robustness of the charge density and supersolid phases for systems away from this regime would require the inclusion of both driving and dissipation in the Hamiltonian \cite{arXiv:1401.5776}.



\begin{thebibliography}{99}
\bibitem{gugu} This onsite energy is chosen to be small enough such that the results are independent of the particular value.
\end{thebibliography}
\end{document}